# Nonlocal Generation of Fano Resonance with No Symmetry Breaking in THz Hybrid Metasurfaces

BOYUAN GE[1], JIAYU FAN[1], KEN QIN[1], XIEXUAN ZHANG[1], HAITAO LI[1], FANG LING[2, 4], XIAOXIAO WU[1, 3, 5]

[1]*Modern Matter Laboratory and Advanced Materials Thrust, The Hong Kong University of Science and Technology (Guangzhou), Guangzhou 511400, China*

[2]*School of Physics and Electronic Engineering, Sichuan University of Science & Engineering, Zigong, 643000, China*

[3]*Low Altitude Systems and Economy Research Institute, The Hong Kong University of Science and Technology (Guangzhou), Guangzhou 511400, China*

[4] *lingfangff@126.com*

[5] *xiaoxiaowu@hkust-gz.edu.cn*

**Abstract:** Fano resonance, arising from the interference between a discrete resonance and a continuum of states, results in sharp and asymmetric line shapes and has significant applications in advanced photonic devices, particularly in sensing, filtering, and nonlinear optics. Nowadays, metasurfaces comprised of engineering microstructures play a crucial role in generation and manipulation of Fano resonance in photonics. However, current metasurfaces dominantly rely on local symmetry breaking of the microstructures to induce Fano resonances, which significant limits their tunability and scalable fabrication for practical applications. To address the challenge, a metal-dielectric hybrid metasurface is demonstrated to achieve nonlocal generation of Fano resonance with no symmetry breaking in the terahertz (THz) band. The Fano resonance, including its existence and peak frequency, is sensitively controlled by the thickness and dielectric constant of the dielectric layer, which is experimentally observed. Our analysis elucidates that the metallic layer with a pair of dumbbell holes leads to the band folding and coupling of guided modes within the dielectric layer. When the thickness or dielectric constant surpasses a critical value, the guided mode resonance falls below the diffraction limit, resulting in a unique nonlocal Fano resonance due to the interaction between the resonance and background transmission facilitated by dumbbell holes. Furthermore, the Fano transmission peak corresponds to an anapole excitation, revealed by multipole calculations. Benefiting from the ability to control the Fano resonance with no symmetry breaking, the proposed hybrid THz metasurface will advance broad applications in the fields of sensors, optical switches, and tunable filters.

## 1. Introduction

Fano resonance, a prominent spectral phenomenon discovered in particle physics [1,2], has garnered substantial attention in photonics owing to its distinctive sharp and asymmetric line shape [3], important for the development of high-Q devices [4,5]. Notably, in periodic metasurfaces, Fano resonance often coexists with Wood's anomalies—both originate from interference between discrete states such as localized resonances or surface waves and continuum states including radiation channels, though Wood's anomalies specifically require momentum matching via periodic modulation [6,7]. In fact, the versatile applications of Fano resonances have spanned across sensing, lasing, spectroscopy, optical communication and many other fields [8-11]. Recently, metasurfaces in the THz frequency range, benefiting from the potential to generate and manipulate Fano resonance for THz applications, has drawn considerable exploration [12-15]. THz metasurfaces are typically composed of metals or dielectrics on dielectric substrates, with patterned microstructures that play the key role [16]. Usually, by introducing local perturbations and defects that break symmetry of the patterned microstructures, the Fano resonance can be generated and excited [17-19]. Meanwhile, due to weak free-space coupling, the decay time of the excited Fano resonance is elongated, which means Fano resonance is able to minimize radiation losses to the greatest extent [20.21]. Thus, the Fano resonance in metasurfaces typically implies optical dark modes with very small radiation loss, which in turn indicates anapole excitation [22-24] and quasi-bound states in the continuum (QBICs) mode [25,26].

The anapole arises from the destructive interference between the excited electric dipole and toroidal dipole, leading to a nonradiative state in the far field [27]. The interference among the multipoles around the emergence of anapole and the background transmission can result in the emergence of Fano resonance peaks [28,29]. To induce anapole and Fano resonance, the metamaterials or metasurfaces usually involve intricately designed three-dimensional structures, and the high-transmission portion of Fano peaks generally corresponds to the frequency where the anapole is excited [30.31]. However, transmission spectrums accompanying anapole arising from planar structures normally exhibits symmetric Lorentzian-shaped peaks [32,33]. Meanwhile, the destructively interfere of multiple radiation channels can generate eigenmodes with infinite lifetime, known as bound states in the continuum (BICs) [34]. By breaking the symmetry that guarantee the exact destructive interference, the BICs generally evolve into QBICs, which often manifest as sharp Fano peaks [35,36]. Thus, most current studies of Fano resonances focus on metasurfaces with asymmetrical or three-dimensional structural units [37-39]. On the other hand, a metasurface that does not rely on local structural symmetry but contingent upon its planar substrate to induce Fano resonance has yet been explored. Moreover, in representative applications such as high-precision sensing, common practices involve placing the analytes of interest in the form of liquids or powders on the metasurface. Potentially, the analytes could damage the fine structures that control the Fano resonance and seriously degrade the precision. Therefore, providing a pathway to generate Fano resonance by only altering the dielectric substrate is highly practical and holds significant potential.

In this paper, we propose and experimentally demonstrate a hybrid THz metasurface that can exhibit Fano resonance with no symmetry breaking. Through the interaction between the guided resonant modes and the background transmission, we can achieve the desired asymmetric spectral response manifested as sharp Fano resonances. Meanwhile, the metasurface utilizes metal structures and dielectric layers to exhibit anapole mode in far-field radiation, and excites QBICs mode when terahertz pulses are incident at slight deviation. The device which has large radius Fano peaks shift for frequency adjustment can utilized for sensing measurements of dielectric constants and substrate thicknesses. This novel method has advantages in device manufacturing, scalability, and design flexibility. We anticipate that this metasurface can be combined with tunable dielectric materials [40] like liquid crystals to realize terahertz optical switches. It can also be designed as an active and tunable device, incorporating active mechanisms for dynamic control of anapole and QBIC modes, thereby opening up new avenues for designing and manufacturing photonic devices as well as driving advancements in THz metasurfaces device applications.

## 2. Experiment and Analysis

### 2.1 The Structure of the THz Hybrid Metasurface

The proposed metal-dielectric THz hybrid metasurface is constructed with two layers, copper (Cu) structure and dielectric substrate, as shown in Fig. 1a. The normal incident THz pulses propagating in the $z$-axis are polarized along the $y$-direction. As shown in Fig. 1b, the Cu layer of 35 μm thickness is structured with periodic dumbbell-shaped holes with lattice consants of $X$ = 800 μm and $Y$ = 400 μm. The dumbbell holes are optimized with radius $R$ = 120 μm and gap distance $G$ = 100 μm. The distance $W$ between the centers of the two circle holes takes the value of 300 μm. Notably, the thickness of substrate $H$ is modified to explore the relationship with the Fano resonance. Fig. 1c presents the metasurface fabricated by the printed circuit board (PCB) manufacturing process and its microscopic image with the thickness of substrate $H$ being 175 μm. The dielectric constant of the substrate is 3.48. Experimentally, we measured the transmission spectrum of the metasurface by the THz time-domain spectrometer (TDS), and the experiment setup is demonstrated in Fig. 1d.

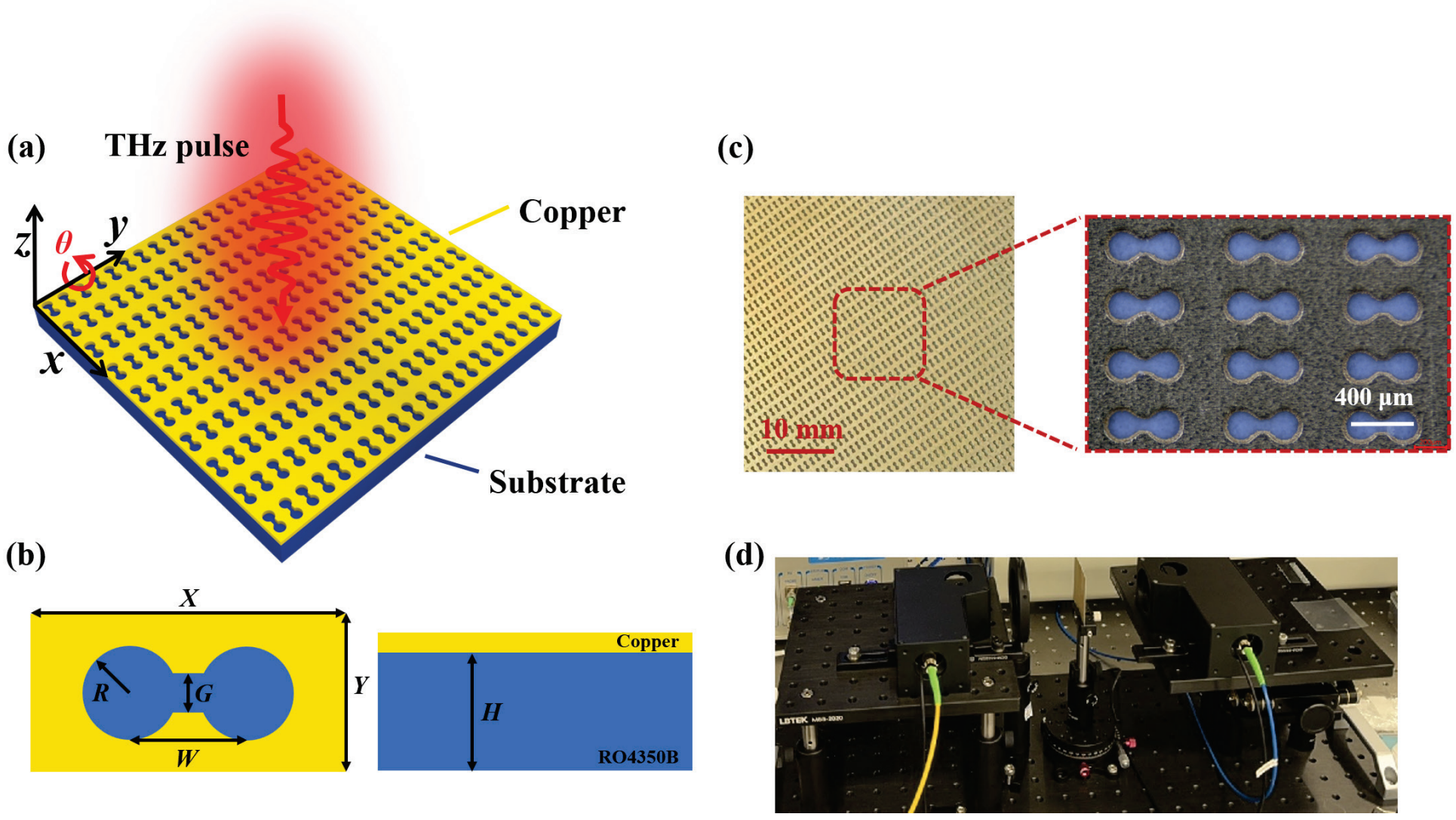

Fig. 1. Structure and experimental design of the THz metasurface. (a) The schematic diagram of the THz hybrid metasurface. (b) The geometric parameters of the structural unit cell. (c) Microscopic image of the manufactured metasurface. (d) A photo of the experimental setup with the THz hybrid metasurface.

### 2.2 Simulation and Measurement of Fano Resonance

The architecture of the dumbbell-shaped metallic layer has been shown to effectively induce an anapole, where the interference between the electric dipole moment (**p**) and the toroidal dipole moment (**t**) results in the cancellation of far-field radiation. This interference produces symmetric Lorentzian transmission peaks [39]. Here, in order to induce asymmetric Fano resonance without disrupting the metallic structure, we incorporate a dielectric layer with an electromagnetic response beneath the metallic layer which perturbs the pristine field, aiming to realize Fano resonance through the combined efforts of both layers.

When THz wave polarized along the $y$-direction is vertically incident on the metasurface, we observed in simulations that with a gradual increase in the thickness of the dielectric substrate (fixed dielectric constant $\varepsilon_r$ = 3.48), a Fano resonance peak suddenly emerges at the frequency of 0.375 THz which is determined by the size of the unit cell as the diffraction limit. The mapped transmissions (Fig. 2a) as a function of frequency and $H$ demonstrates the characteristics of the sudden appearance and gradual shifting of the Fano resonance peak frequency as the substrate thickness $H$ varies from 100 μm to 200 μm. Starting from the thickness of $H$ = 130 μm, where the Fano peak initially develops, we extracted the frequencies of Fano peaks at an interval of 10 μm and plotted them in Fig. 2b. As can be seen, the Fano resonance peak is shifted towards lower frequencies from 0.375 THz to 0.335 THz when the substrate thickness $H$ increases from 130 μm to 200 μm.

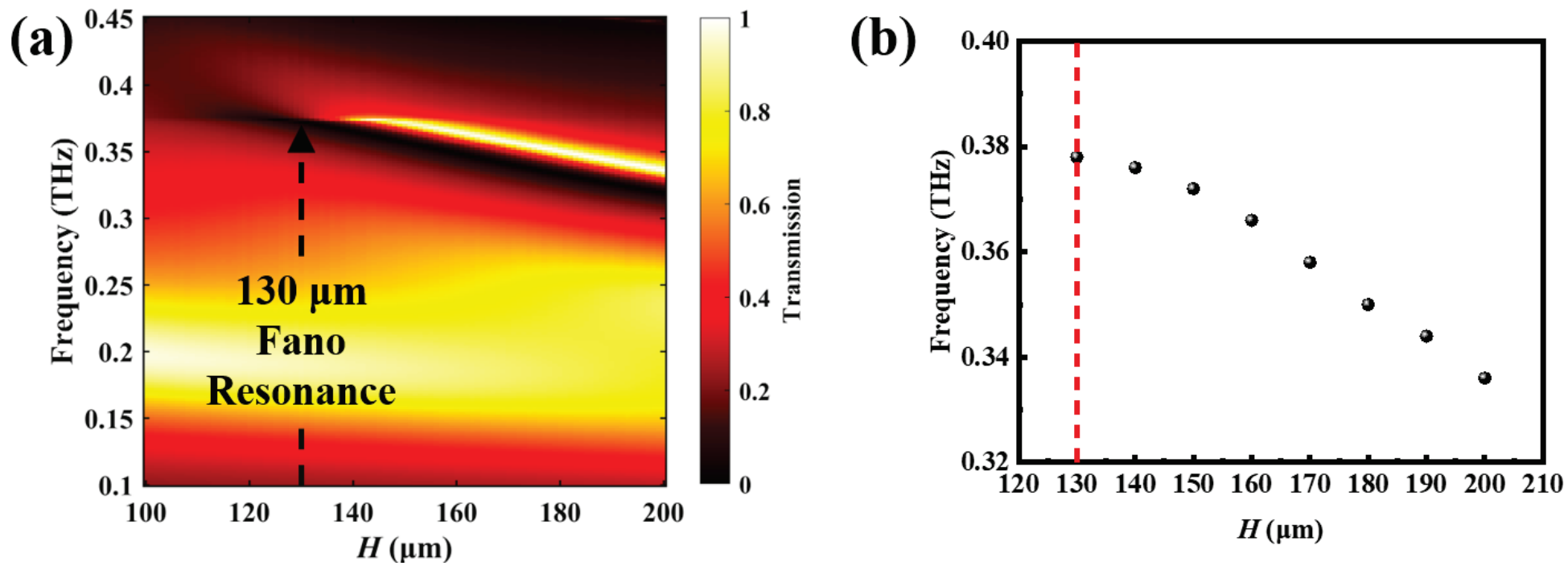

Fig. 2. Emergence of Fano resonance. (a) The transmission spectrum mapped versus the thickness of the dielectric layer $H$. A sharp Fano resonance starts to emerge when $H \geqslant 130$ μm. (b) The frequency of the Fano peak for every 10 μm increase in substrate thickness $H$. The dashed line denotes the critical thickness 130 μm that Fano resonance emerges.

We produced two samples with dielectric layer thicknesses of 110 μm and 175 μm to experimentally compare the emergence or not of the Fano resonance. In the experimental measurements conducted via TDS, we distinctly observe a pronounced Fano peak at 0.354 THz (Fig. 3a), accompanied by a near-zero transmission at 0.336 THz. These key signatures of the Fano resonance closely align with the frequencies predicted from the simulated results. However, when the dielectric layer thickness is 110 μm (Fig. 3b), there is no Fano resonance observed in the transmission spectrum, in both simulation and experimental results. It is worth mentioning that the broad transmission centered around 0.2 THz is predominantly generated independently by the dumbbell metallic layer. In Fig. 3c, we fitted the Fano transmission resonance generated by the sample with the thickness of 175 μm using Equation (1), where $t(\omega)$ represents the transmission, $t_0$ represents the background transmission, $\omega_0$ is the resonance center angular frequency, $\omega$ is the incident angular frequency, and $\gamma_0$ is the resonance linewidth [3]. The asymmetry factor of the Fano resonance $q \approx 2.8$ is obtained, and the quality factor Q ≈ 40.

$$t(\omega) = t_0 \frac{|1 + (\omega - \omega_0)/(q\gamma_0)|}{\sqrt{1 + [(\omega - \omega_0)/\gamma_0]^2}} \tag{1}$$

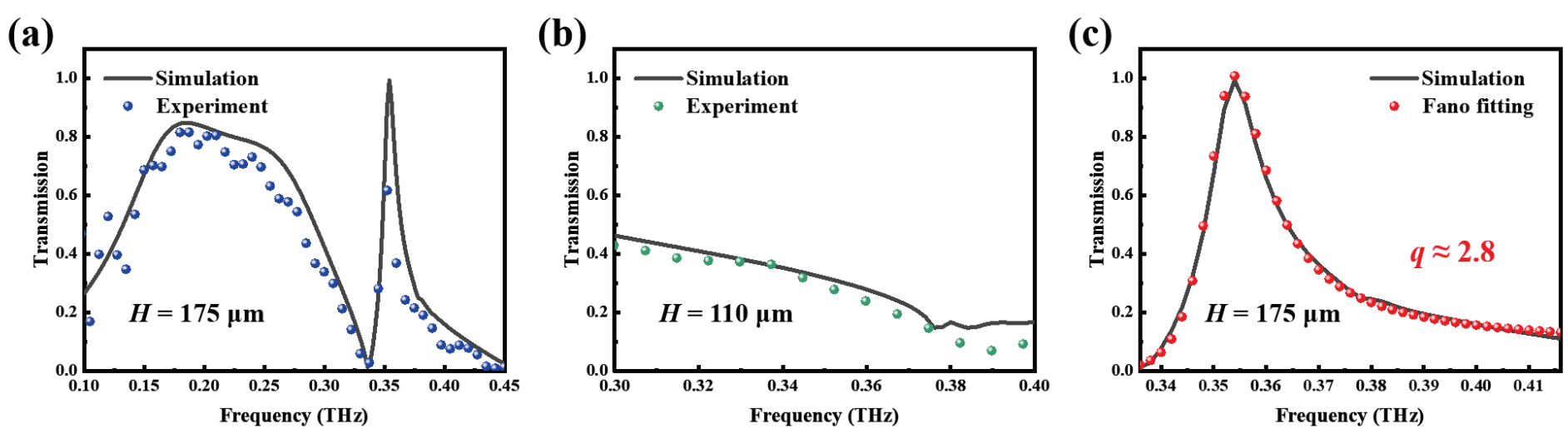

Fig. 3. Transmission spectra of two samples with different thicknesses. (a) Simulated and measured transmission spectra for the sample with the $H = 175$ μm dielectric substrate thickness. (b) Simulated and measured transmission spectra for the sample with the $H = 110$ μm dielectric substrate thickness. (c) Fitting of the Fano resonance with Equation (1) for the sample with the $H = 175$ μm dielectric substrate thickness.

The experimental, simulated, and fitted results of the Fano resonance exhibit a good match, with the transmission amplitude exceeding 0.6 in the experimental measurements which indicates the practical application and capability of our designed metasurface in the THz band. The experimental transmission of the Fano peak is capped mainly due to the limited frequency resolution (2.5 GHz) of our THz-TDS system, which restricts the ability to fully resolve sharp features of the resonance.

### *2.3 Multipole Radiation and Anapole*

Before delving further into investigating the cause of the Fano resonance, studying the scattering properties of metasurfaces is essential. Multipolar decomposition analyses are commonly used to analyze the scattering power of multipole moments in the far field. In Fig. 4a, the multipole decomposition analysis illustrates the contributions of different multipole moments - electric dipole P, magnetic dipole M, toroidal dipole T, electric quadrupole $Q_e$, and magnetic quadrupole $Q_m$ to the far-field radiation intensities of the 175 μm dielectric substrate metamaterials. At the Fano resonance peak frequency of 0.354 THz, the amplitudes of P and T nearly overlap, indicating comparable strengths, while M, $Q_e$, and $Q_m$ exhibit weaker intensities. Evidently, P and T dominate the multipolar radiation at this Fano peak frequency. For the non-radiating anapole, the far-field radiation intensity is given by [41]:

$$I = \frac{k_0^2}{6\pi{\epsilon_0}^2} |\mathbf{p} - ik_0\mathbf{t}|^2 = 0 \qquad (2)$$

where the vector electric dipole moment $\mathbf{p}$ and toroidal dipole moment $\mathbf{t}$ satisfy $\mathbf{p} = ik_0\mathbf{t}$, resulting in the far-field radiation vanishing due to interference cancellation. We calculated the radiation intensity $I$, for three thicknesses of 110 μm, 130 μm, and 175 μm. When the dielectric material is thin, the anapole provided by the metallic dumbbell layer itself predominates, resulting in $I = 0$ near the range of 0.15 THz to 0.2 THz (Fig. 4b). As the thickness of the dielectric substrate increases, the intensity turning point towards 0 occurs at higher frequencies as the metallic layer and dielectric layer interact. For the 175 μm sample we fabricated, the radiation intensity is nearly 0 at 0.354 THz (Fig. 4b) which can be attributed to the presence of trace amounts of multipole radiation, generating a new frequency distinct from the anapole provided solely by the metal dumbbell metamaterial surface.

To further understand the generation mechanism of anapole in hybrid metasurfaces, we conducted a study on the near-field electromagnetic distribution of metasurfaces at the Fano resonance frequency of 0.354THz. In Fig. 4c, the electric field and current distribution show a concentration of the electric field near the gap. From the *xy*-plane perspective, the current arrows are predominantly along the -*y* direction due to the action of the electric dipole moment in the metal layer. More details on the electromagnetic field distribution can be found in Supplement 1 Fig. S1. Regarding the dielectric layer, the *xz*-plane in Fig. 4d reveals a swirling magnetic field distribution, indicating the excitation of a toroidal dipole moment along the +*y* direction in the hybrid metasurface. The excitation of this swirling magnetic field is closely related to the thickness of the substrate layer, with the majority existing within the dielectric substrate. At the Fano resonance frequency of 0.354 THz, this swirling magnetic field reaches its peak. (For magnetic field distributions related to thickness and frequency, refer to Supplement 1 Fig. S2) Hence, the electric dipole moment dominated by the metal layer and the toroidal dipole moment excited in the dielectric layer counteract each other at the Fano resonance frequency, leading to interference cancellation that satisfies the anapole condition.

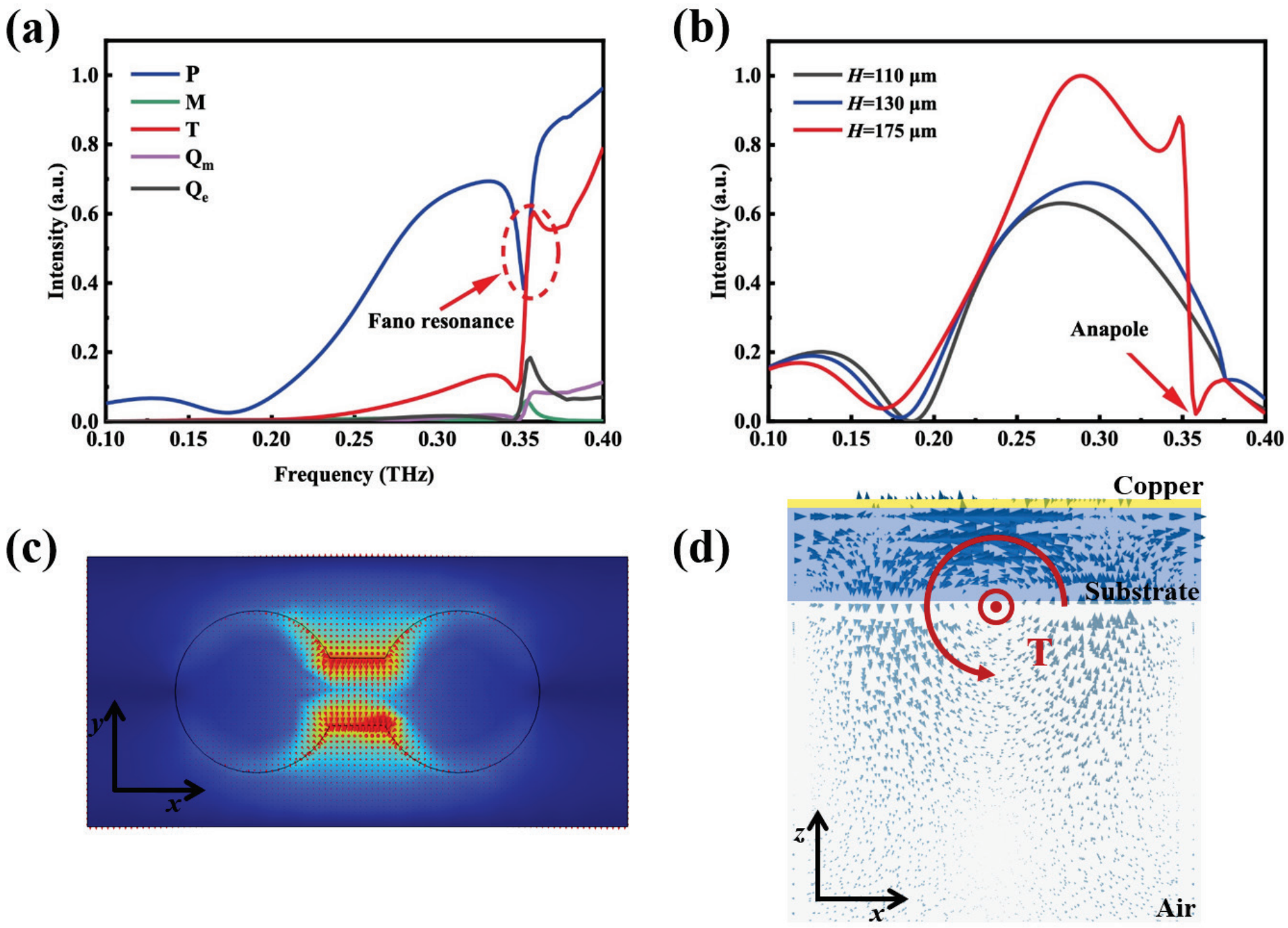

Fig. 4. Multipole analysis of the hyrid metasurface. (a) Multipolar contribution of electric dipole P, magnetic dipole M, toroidal dipole T, electric quadrupole $Q_e$, and magnetic quadrupole $Q_m$ to radiation intensities. Scattering power of different multipoles calculated by multipolar decomposition. (b) The total radiation intensity of electric dipole moment **p** and toroidal dipole moment **t**. (c) The electric field and current distribution on metasurface. (d) Representation of swirling magnetic field distribution in the *xz*-plane.

### *2.4 Guided-Mode Resonances in Hybrid THz Metasurfaces*

Here, we proceed to confirm the decisive role of the dielectric substrate positioned beneath the metallic layer on Fano resonance. In fact, the dielectric substrate act as a slab waveguide, while the periodic metallic structure serves as a diffraction grating, thereby leading to the formation of the nonlocal guided-mode resonances (GMRs) [42]. The high transmission generated by this hybrid structure differs fundamentally from the multiple interference effects of multilayer metasurfaces, which involves no guided modes [43]. We computed the band diagrams (Fig. 5a) for dielectric substrates with thicknesses of 100 μm, 175 μm, and 300 μm to further analyze the role and mechanism of the GMRs. The gray region delineates the segment enclosed by the light line. Upon inspection, the 100 μm thick substrate (depicted in red) entirely resides within the gray region. As the thickness escalates to 175 μm, a green band emanates from the gray region, extending beyond the light line. In an ideal scenario of an infinite and homogeneous dielectric substrate, band folding is non-existent. Nevertheless, due to the periodic nature of the upper metallic structure, we folded the green band at the M point. Our analysis revealed its proximity to the Γ point at approximately 0.354 THz, aligning with the resonance position of the observed Fano peak in the 175 μm configuration. Upon reaching the thickness of 300 μm, two blue bands manifest beyond the light line, elucidating the emergence of the second Fano peak in subsequent experiments. We propose that the interaction between the leaky mode and the guided mode in the dielectric layer gives rise to guided-mode resonance. In the absence of a dielectric slab, when a THz wave polarized in the *y*-direction is incident upon the periodic structure at $Y$ = 400 μm, the diffraction-limited frequency is 0.75 THz. However, the Fano resonance occurs at frequency of 0.375

THz, arising from the generation of the guided mode that propagates in the *x*-direction along $X = 800$ μm and manifests as a transverse wave. The upper metal layer provides an electric field polarized in the *y*-direction, which propagates along the *x*-direction within the dielectric layer. When the thickness of the dielectric slab is insufficient, the red band remains within the light line and disjoints at the Γ point (Fig. 5a). As the thickness increases, for the band at Γ point with frequencies below 0.375 THz, the slab waveguide thickness of 175 μm first occurs at 0.283 THz. However, the guided wave mode is distributed in the x-direction, which is incompatible with the incident electric field at this frequency, as shown in Fig. 5b. When the guided mode outside the light line folds back to Γ, the frequency is 0.354 THz. At this point, the slab waveguide sufficiently supports the existence of guided modes polarized in the y-direction, allowing the y-polarized incident wave provided by the metal layer to excite this guided mode, as depicted in Fig. 5c. Due to the generation of GMRs, the bands outside the light line fold into the Γ point, resulting in the emergence of leaky modes. The remaining momentum is emitted in the *z*-direction, leading to the transmission of Fano resonances.

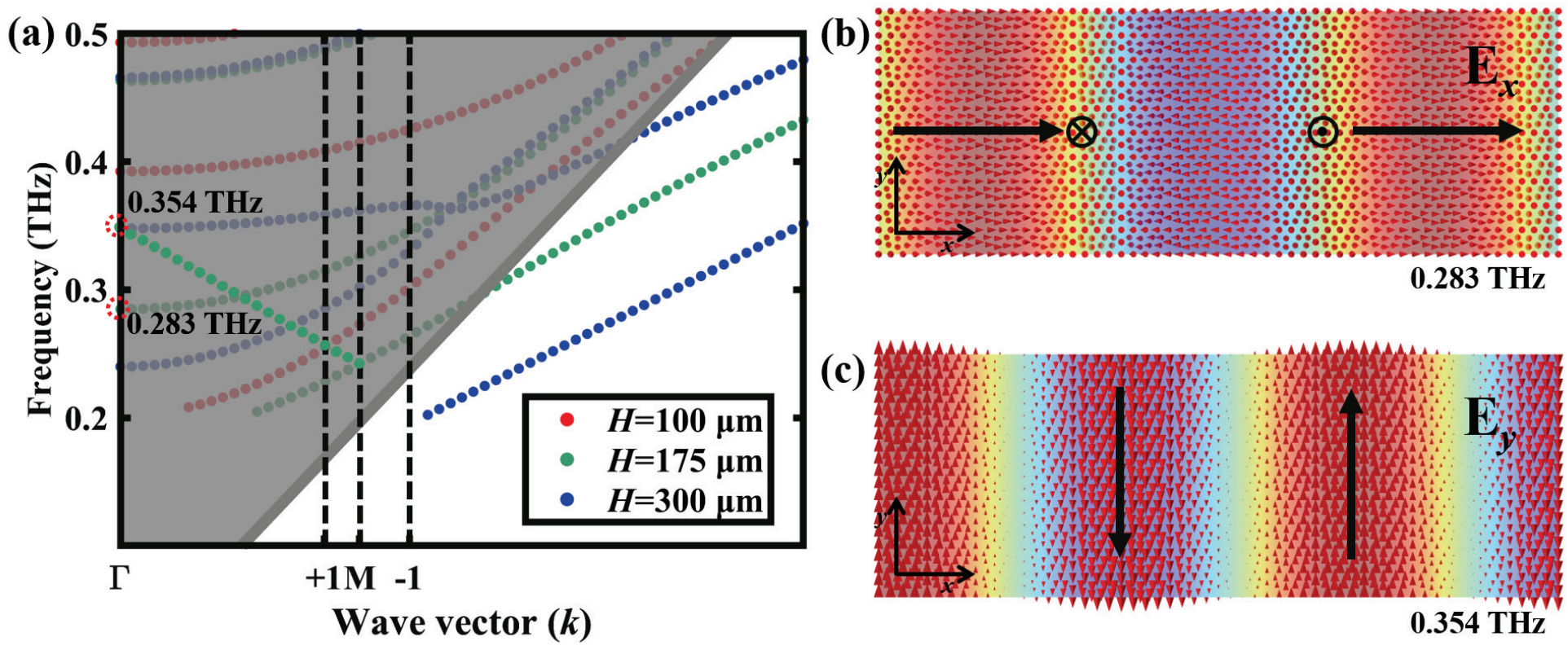

Fig. 5. Bands and non-local GMRs. (a) Band diagrams for three different thicknesses (100 μm, 175 μm and 300 μm) of dielectric substrates. (b) and (c) show the electric field and current distribution of the 175 μm thick dielectric slab waveguide at frequencies of 0.283 THz and 0.354 THz, respectively.

To demonstrate the presence of GMRs in this structure, we conducted simulations and experiments by varying the incident angles [44]. At oblique incidence as the metasurface is rotated along the *y* axis by angle $\theta$ (denoted in Fig. 1a), two distinct frequency-detuned GMRs are excited generally, and the frequency difference between these GMRs increases with the angle of incidence. Corresponding to simulations (Fig. 6a) and experiments (Fig. 6c) with the 110 μm thick dielectric substrate, we observe that there are no peaks or variations in the transmittance spectra within the range of 0.3 THz to 0.4 THz, even when the incident angle deviates from normal incidence. However, the Anapole transmission peak at 0.2 THz remains unaffected by changes in the incident angle. In contrast, Fig. 6b and Fig. 6d present the simulation and experimental results for the 175 μm thick dielectric substrate. When the incident angle is slightly adjusted, the original Fano peak splits into two peaks: one shifting towards higher frequencies (-1) and the other towards lower frequencies (+1). As the tilt angle increases such that the frequency of -1 GMR exceeds the diffraction limit of the structural period (0.375 THz), the -1 resonance peak fades away. Notably, at very small angles of incidence, destructive interference occurs between these backward-propagating GMRs, leading to the emergence of QBICs in the continuum. The mode, initially bound states in the continuum safeguarded by symmetry under normal incidence, transforms into QBICs stimulated by plane waves incident at an angle. This explains why no transmission peak is present at the initial position of the +1 GMR. Due to the presence of hybridization within the resonance peak, the evolution is not as perfect as in classical GMRs. However, the excellent agreement between simulations and experiments sufficiently demonstrates presence of the nonlocal GMRs in the metasurface, which plays the key role in the formation and modulation of Fano peaks.

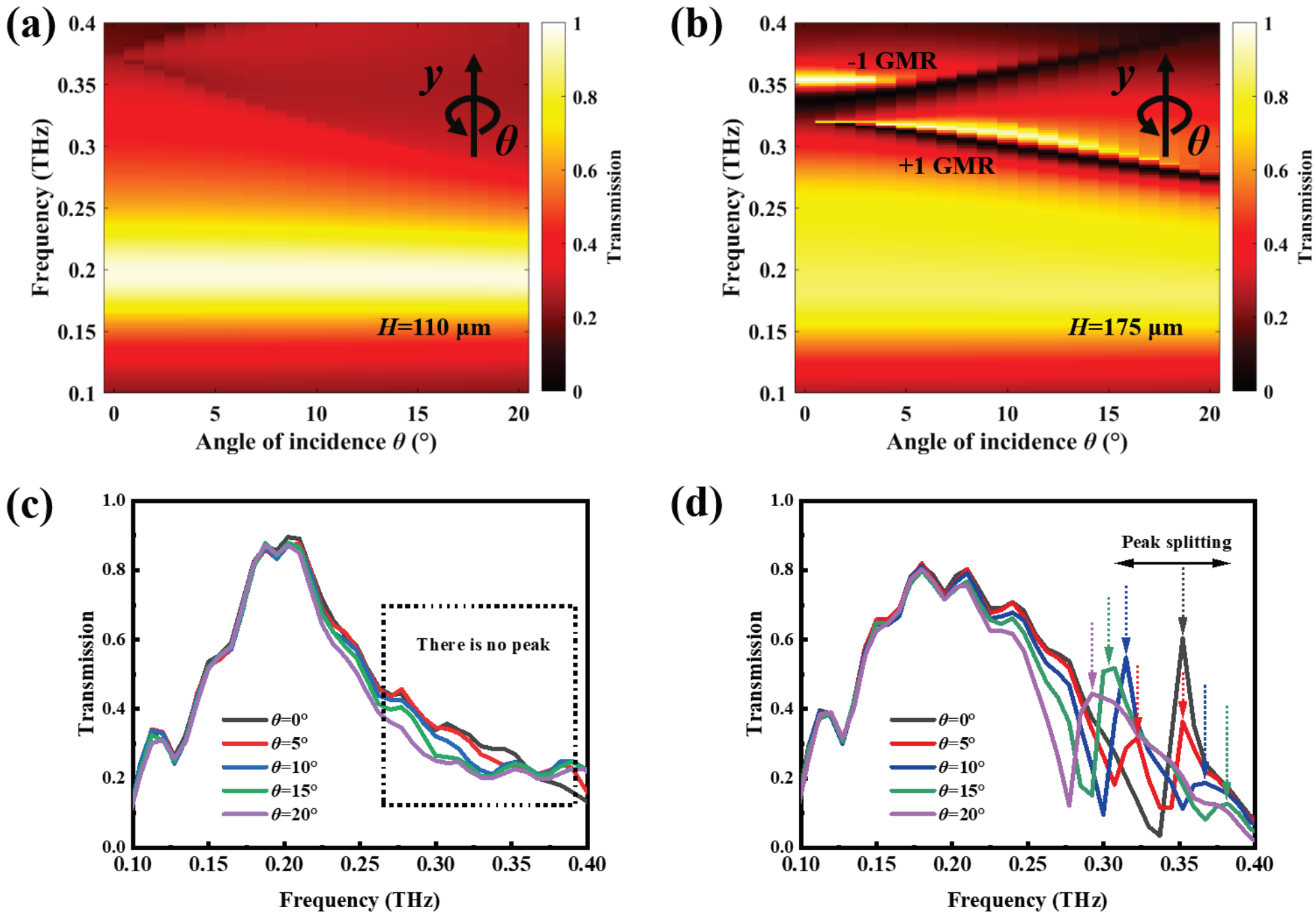

Fig. 6. Transmittance spectra at different incident angles in simulations and experiments. The angle $\theta$ rotates about the $y$-axis in the coordinate system from 0° to 20° (see also Fig. 1a). (a) and (c) Transmittance spectra in simulations and experiments for the 110 μm thick dielectric substrate as the incident angle is varied. (b) and (d) Transmittance spectra in simulations and experiments for the 175 μm thick dielectric substrate as the incident angle is varied.

### *2.5 Sensing Applications of Hybrid THz Metasurfaces*

Terahertz metasurfaces with sharp Fano resonances hold promise for high-sensitivity sensing applications. Leveraging the relationship between Fano resonances of metasurfaces and dielectric thickness, we have envisioned and realized flexible sensors. Diverging from conventional terahertz metasurfaces placing the sample on top of the structure, we conduct measurements from the bottom of metasurfaces to analyze substances, enabling straightforward and non-destructive applications for measuring dielectric material ultra-thin thickness and permittivity by monitoring the movement of the Fano peak. This approach can be extended to various sensing domains based on specific application requirements. We initially conducted tests on polyethylene terephthalate (PET) films of six different thicknesses. The simulation results and experimental results are depicted in Fig. 7a and Fig. 7b. As illustrated in Fig. 7c, we adhered PET films to the bottom, with Fig. 7d showcasing the deviation of the Fano peak frequency between the experimental and simulated data. With increasing in film thickness, the Fano peak shifts towards lower frequencies, enabling the measurement of dielectric film thickness by comparing new peaks with the original frequency.

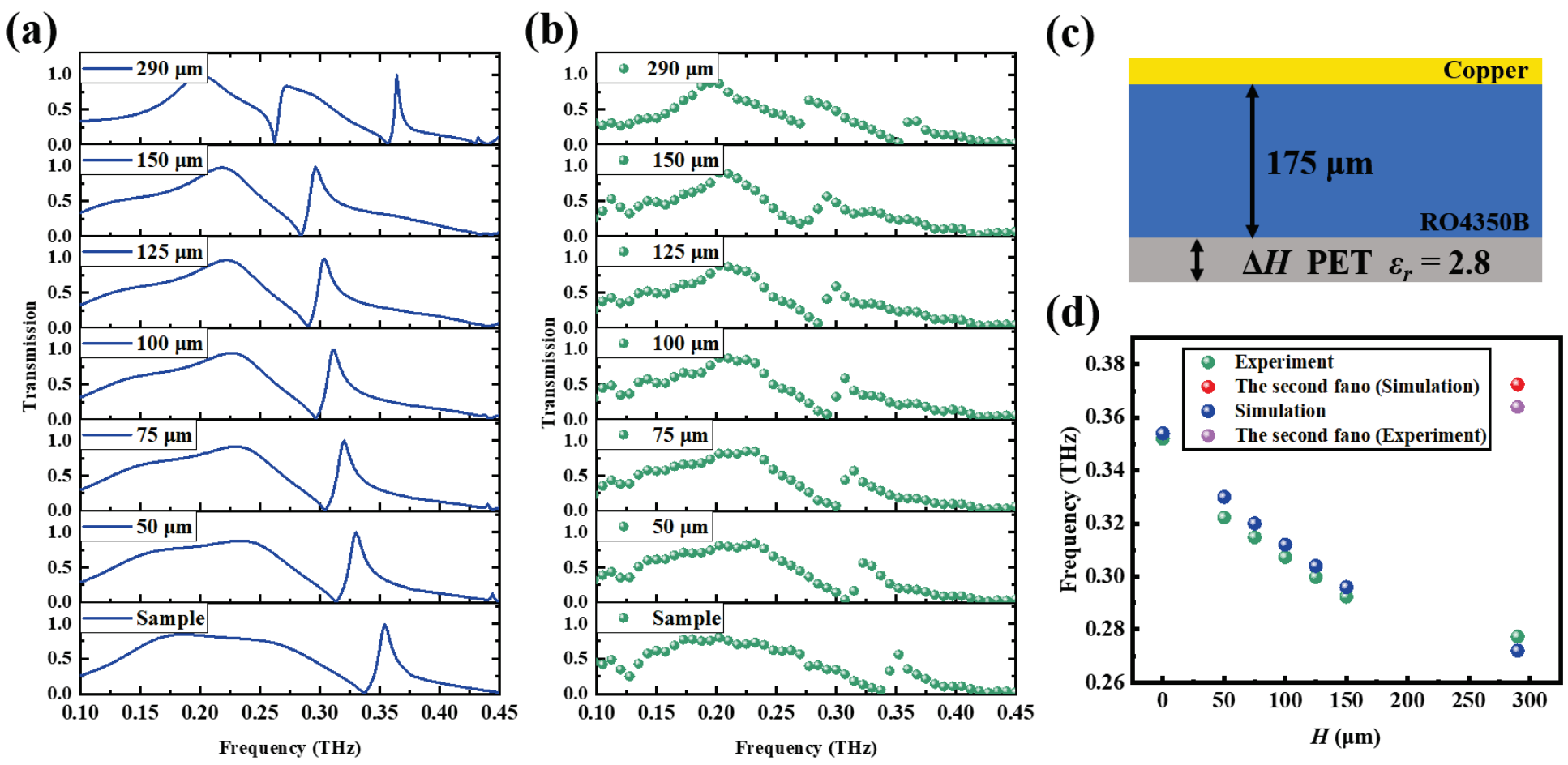

Fig. 7. PET of different thickness. (a) and (b) Simulation and experimental results of the transmittance spectra as the thickness of the PET film increases. The appearance of the second Fano peak confirms the presence of the second band in the dielectric substrate, as mentioned in Fig. 5a. (c) Schematic diagram of the measurement method for the film, dielectric constants ($\varepsilon_r$ = 2.8). (d) The comparison between measurement and simulation errors demonstrates excellent sensing performance.

Next, we conducted tests on three different films with same thickness but different dielectric constants: polyimide (PI), PET, and polydimethylsiloxane (PDMS), to validate applications of metasurfaces in measuring dielectric constants. Fig.8 a-c display the thickness - 50 μm and dielectric constants ($\varepsilon_r$) of the three films. Fig.8 a-c present the simulation and measurement results, showing that for the same thickness, the larger dielectric constants lead to greater shifts in the Fano peak frequency. The error of approximately 0.007 THz is indicated in Fig. 8f. In Supplement 1 Fig. S4, we investigate the critical dielectric constant $\varepsilon_r$ and thickness $H$ required to generate Fano resonances, and find that the critical value of $\varepsilon_r$ is inversely proportional to $H^{1.32}$.

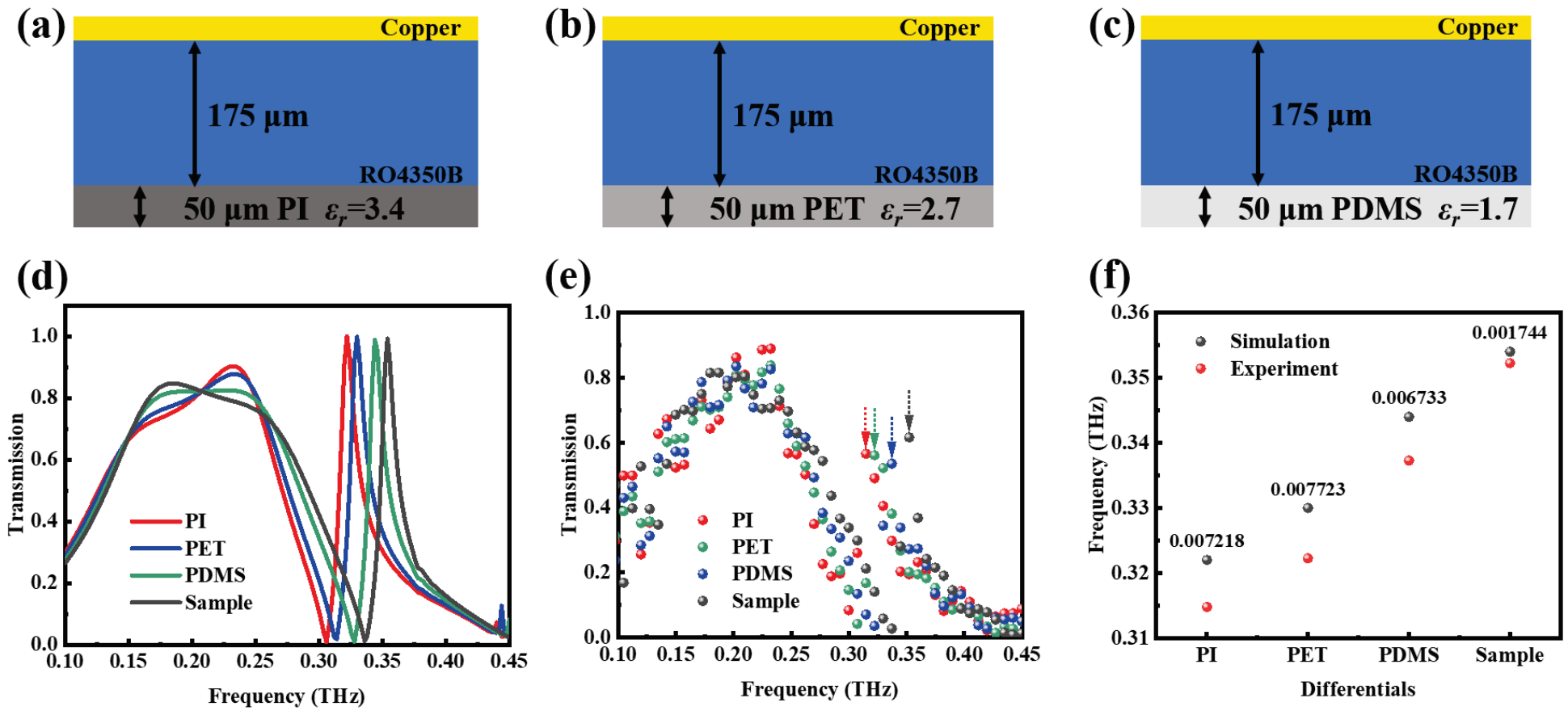

Fig. 8. Three films with different dielectric constants. (a)-(c) represent schematic diagrams of the structures during the measurements. (d) Simulation results of the three different films and the original Fano peak. (e) Experimental measurements of the Fano peak shifts for the three different films compared to the original Fano peak without any sample. (f) Error display of the simulation and experimental results.

## 3. Discussion and Conclusion

In summary, this study introduces a terahertz metasurface featuring Fano resonances through the hybridization of metal and dielectric materials, all without disrupting symmetry. The underlying mechanism is revealed through electromagnetic characteristics in near and far fields, along with band structure analysis. The introduction of dielectric slab waveguides demonstrates terahertz guided mode resonances, accompanied by multipolar radiation characteristics of anapole modes in the far field. Excellent consistency between simulations and experiments confirms the feasibility of this approach. Regarding applications, the sensitivity of metasurfaces to dielectric materials enables the measurement of films with varying dielectric constants and thicknesses. With the utilization of this composite metasurface design, we are poised to drive the development and application of terahertz metasurfaces in dynamic tunable devices.

**Funding.** This work is supported by National Natural Science Foundation of China (No. 12304348), Guangdong Basic and Applied Basic Research Foundation (No. 2025A1515011470), Guangdong University Featured Innovation Program Project (2024KTSCX036), Guangzhou-HKUST(GZ) Joint Funding Program (2025A03J3783), Guangzhou Municipal Science and Technology Project (No. 2024A04J4351). F. Ling acknowledges support by the Scientific Research and Innovation Team Program of Sichuan University of Science and Engineering (SUSE652B004). X. Wu acknowledges support from the Modern Matter Laboratory in HKUST(GZ).

**Disclosures.** The authors declare no conflicts of interest.

**Data availability.** Data underlying the results presented in this paper are not publicly available at this time but may be obtained from the corresponding author upon reasonable request.

**Supplemental document.** See Supplement 1 for supporting content.